\documentclass[doublecol]{epl2} 
\usepackage{amssymb,amsmath}
\newcommand{\eeta}{{\boldsymbol{\eta}}}

\title{Out of equilibrium thermal Casimir effect between Brownian conducting plates}
\shorttitle{Thermal Casimir effect between Brownian conductors}

\author{B.-S. Lu\inst{1,2} \and D.S. Dean \inst{3} \and R. Podgornik\inst{1,2}}
\shortauthor{B-S Lu \etal}

\institute{                    
  \inst{1}Department of Physics, Faculty of Mathematics and Physics, University of Ljubljana, Jadranska ulica 19, SI-1000 Ljubljana, Slovenia \\
  \inst{2}Department of Theoretical Physics, J. Stefan Institute, 1000 Ljubljana, Slovenia \\
  \inst{3} Universit\'e de  Bordeaux and CNRS, Laboratoire Ondes et
Mati\`ere d'Aquitaine (LOMA), UMR 5798, F-33400 Talence, France
}

\abstract{We study the thermal fluctuation induced interactions between two surfaces containing Brownian charges which are held at different temperatures. Using a dynamical form of  Debye-H\"uckel theory implemented within the stochastic equation for the density of mobile Brownian charges, we derive expressions for the average force between the two surfaces as well as its variance. The latter is found both for the normal, of finite mean, as well as the lateral force, of zero mean, between the surfaces. }

\begin{document}

\maketitle

\section{Introduction}
Recently there has been intense interest in the out of equilibrium Casimir and/or van der Waals interactions between objects held at different temperatures \cite{dor98,ant05,ant08,bim09,kru11a,kru11b,mes11,rod11}. The analysis of such systems, taking into account both quantum and finite temperature effects, is carried out using the {\em stochastic electrodynamics} formalism of Rytov \cite{ryt89}. This is a very general formalism based on Maxwell's equations in the presence of random currents which satisfy the quantum fluctuation  dissipation theorem. The correlation function of currents in the same body, assumed to be held at fixed temperature, obeys a quantum fluctuation dissipation theorem with the temperature of the body in question. The electrical properties of the media in the Rytov formalism are encoded in their dielectric functions. 

Here we will study  a classical problem from a more microscopic, and statistical mechanics, point of view. The basic model consists of two plates in which a single species of charge carrier, with a uniform electroneutralizing background charge (a so called {\em jellium} model), is described by an over-damped Langevin equation with an applied force generated by a direct Coulomb interaction with the charges in the same and opposing plates. In this formalism the dynamical temperature of each plate is well defined and appears as the ratio of the Langevin noise (diffusion constant) to the mobility relating the drift to the applied force (the local Einstein relation).  Within this approximation we compute the average thermal Casimir   
force between the two plates and are able to give explicit formulas for the near and far field forces in terms of the microscopic parameters of the problem. The far field results are found to have the same dependence on the plate separation $L$ as results derived within the Rytov formalism \cite{dor98,ant05,ant08,bim09,kru11a,kru11b,mes11,rod11}. In addition 
the model here allows us to compute the variance of both the normal and lateral forces
in this out of equilibrium context and  to our knowledge these are the first results on force 
fluctuation statistics for the  non-equilibrium thermal Casimir effect. 
 
\section{The model}
\begin{figure}
\centerline{\includegraphics[width=8cm]{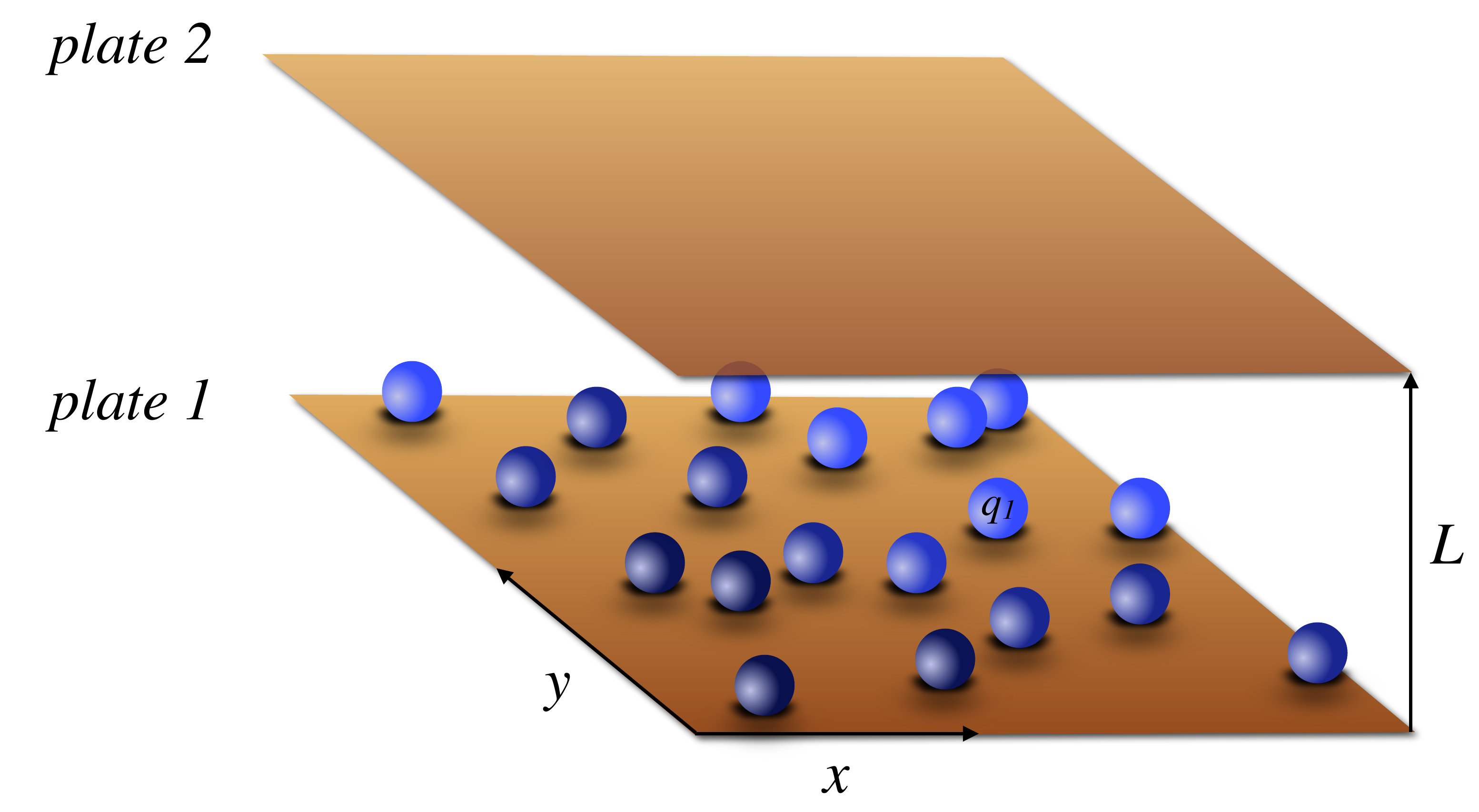}}
\caption{Schematic of the model, two plates 1 and 2 containing Brownian charges of charge $q_1$ and $q_2$ (not shown) respectively. The separation between the plates is $L$ and the in plane $(x,y)$ coordinate axes is shown for plate 1. All charges interact via the direct non-retarded three dimensional Coulomb interaction.}
\label{schematic}
\end{figure}

We will consider a system of two parallel planes of area $S$ separated by a distance $L$
as shown in fig.~(\ref{schematic}). Each plane $\alpha$ contains a uniform background charge density (per unit area) $\overline \rho_{c\alpha}$ along with a species of mobile ion $\alpha$, of charge $q_\alpha$ and diffusion constant $D_\alpha$, obeying the Langevin equation
\begin{equation}
{d{\bf X}\over dt} = \beta_\alpha D_\alpha q_{\alpha}{\bf E} + \sqrt{2D_\alpha}\eeta_{\alpha}.
\end{equation}
Above $\eeta_{\alpha}$ is zero mean Gaussian white noise with correlation function
$\langle \eta_{\alpha i}(t) \eta_{\alpha j}(t') \rangle=\delta_{ij}\delta(t-t')$ and $\bf E$ is the local electric field in the plane which is generated by the electric charge distributions in both planes. The temperature of the plate $\alpha$ is $T_\alpha  =1/\beta_\alpha$ and thus the electrical mobility of the species $\alpha$ is given by $\mu_\alpha = \beta_\alpha D_\alpha$ from  the Einstein relation. The statics of the above model within the Debye-H\"uckel approximation has been studied by a number of authors to understand how charge fluctuations in neutral colloidal systems give rise to attractive interactions \cite{lev99}
as well as to better understand the high temperature limit of the Casimir force between conductors \cite{jan05,bue05}. More recently \cite{dea14} a stochastic density functional approach was applied to the same system in order to see how the thermal Casimir force evolves in time between two plates having initially no charge fluctuations. Interestingly the relaxation of the thermal Casimir force to its equilibrium value turns out to be very different from that for a model dielectric based on polarisable dipole dynamics \cite{dea12}, even if both models can have identical equilibrium Casimir forces in the appropriate limits. 

Within each plate $\alpha$ the in plane component of the electric field can be expressed in terms of the density operator  of each charged species defined by
\begin{equation}
\rho_\alpha({\bf x}) = \sum_{i=1}^N \delta({\bf x} -{\bf X}_{\alpha i})
\end{equation}
and is given in the plane $\alpha/\beta$ by
\begin{align}
{\bf E}_{\alpha/\beta} &= -{1\over \epsilon} \int_S d{\bf x'}\ \left[ {\nabla}_{||} G({\bf x}-{\bf x'},0)q_{\alpha/\beta}\rho_{\alpha/\beta}({\bf x}')\right.\nonumber\\
 &\left.+ {\nabla}_{||} G({\bf x}-{\bf x'},L)q_{\beta/\alpha}\rho_{\beta/\alpha}({\bf x}') \right],\label{eE}
\end{align}
where $\nabla_{||}$ denotes the gradient operator in the $(x,y)$ planes containing the charges. Here $\epsilon$ is the dielectric constant of the system which is assumed to be constant and $G$ is the Green's function obeying
\begin{equation}
[\nabla_{||}^2  +{\partial ^2 \over \partial z^2}] G({\bf x}-{\bf x}', z-z')=-\delta({\bf x}-{\bf x}')\delta(z-z').
\end{equation}
 We also note that in Eq. (\ref{eE}) the uniform background charge in both planes do not contribute to the in plane electric field. It is useful to define the fluctuations of the overall
density (per unit area) of species $\alpha$ with respect to the mean value $\overline {\rho}_\alpha$ via 
\begin{equation}
\rho_{\alpha}({\bf x}) = \overline {\rho}_\alpha + n_\alpha({\bf x}).
\end{equation}
In terms of the field $n_\alpha({\bf x})$ one can write the force acting perpendicular
to the plates as
\begin{equation}
f_{\perp} = -{q_1 q_2\over \epsilon} \iint_S d{\bf x} d{\bf x'} n_1({\bf x})n_2({\bf x'}){\partial\over \partial L}G({\bf x}-{\bf x'},L).\label{fp}
\end{equation} 
The $i-th$ component  force parallel to the plates, which will be on average zero, is given by
\begin{equation}
f_{\parallel,i}=  -{q_1q_2\over \epsilon} \iint_S d{\bf x} d{\bf x'}\ n_1({\bf x})n_2({\bf x'}) \nabla_i G({\bf x}-{\bf x'},L).\label{fl}
\end{equation}
The density fields $\rho_\alpha$ in each plate can be shown to evolve according to the stochastic partial 
differential equation \cite{kaw94,dea96}
\begin{align}
{\partial \rho_\alpha({\bf x},t) \over \partial t} &= D_\alpha\nabla_{||} \cdot\left[ \nabla_{||} \rho_\alpha({\bf x},t)
-\beta_\alpha q_\alpha{\bf E}_\alpha \rho_\alpha({\bf x},t)\right] \nonumber \\&+ \nabla_{||}\cdot[ \sqrt{2 D_\alpha
\rho_{\alpha}({\bf x},t)} \eeta_\alpha({\bf x},t)].\label{edyn}
\end{align}  
In the above, the noise $\eeta_\alpha({\bf x},t)$ is a spatio-temporal white noise vector field of mean zero and with correlation function 
\begin{equation}
\langle \eta_{\alpha i}({\bf x},t)\eta_{\beta j}({\bf x}',t)\rangle = \delta_{\alpha\beta}\delta_{ij}\delta(t-t')\delta({\bf x}-{\bf x}').
\end{equation}
\section{Dynamical Debye-H\"uckel Approximation}
As it stands eq.~(\ref{edyn}) is a non-linear (as the electric field depends on the density) partial differential equation with a noise term that depends on the local density field and is not amenable to an exact mathematical analysis. However, one can linearise the equation about the mean value of the density to find a dynamical theory which 
yields the Debye-H\"uckel high temperature/weak coupling theory as its static limit \cite{dea14}. The equation is expanded to first order in $n_\alpha$ in the deterministic part and to zeroth order in the noise part  (the first order correction to the noise is on average zero so this is compatible with expanding the deterministic part to first order). The resulting   linear equation in Fourier ({\bf Q}) space reads, using the Einstein summation convention on repeated indices, 
\begin{equation}
\partial_t \tilde n_\alpha({\bf Q},t) = -M({\bf Q})_{\alpha\beta} \tilde n_\beta({\bf Q},t) + \xi_\alpha({\bf Q},t),
\label{eq:langevinesque}
\end{equation}
where the noise term $\xi_\alpha({\bf Q},t)$ has correlation function
\begin{equation}
\langle \xi_{\alpha}({\bf P},t) \xi_{\beta}({\bf Q},t') \rangle = 2(2\pi)^2 \delta({\bf P}+{\bf Q}) \delta(t-t') R_{\alpha\beta}({\bf Q}),
\end{equation}
where 
\begin{equation}
R_{\alpha\beta}({\bf Q}) =\delta_{\alpha\beta} D_\alpha \bar{\rho}_\alpha Q^2
\end{equation}
and 
\begin{align}
&M({\bf Q}) =\nonumber \\  &Q^2 \begin{pmatrix}
      D_1 + q_1^2 D_1 \beta_1 \bar{\rho}_1 \tilde G({\bf Q}.0) & q_1 q_2 D_1 \beta_1 \bar{\rho}_1 \tilde G({\bf Q},L)          \\
      q_1 q_2 D_2 \beta_2 \bar{\rho}_2 \tilde G({\bf Q},L) & D_2 + q_2^2 D_2 \beta_2 \bar{\rho}_2 \tilde G({\bf Q},0)
\end{pmatrix},
\end{align}
where $\tilde G({\bf Q},z)=e^{-Qz}/2Q$ is the Fourier transform of $G$ taken in the $(x,y)$ plane. From this one can show \cite{zwa01} that the steady state correlation function defined by
\begin{equation}
\langle \tilde n_\alpha({\bf Q})\tilde n_\beta({\bf P})\rangle = (2\pi)^2 \delta({\bf Q}+{\bf P})\tilde C_{\alpha\beta}({\bf Q})
\end{equation}
obeys
\begin{equation}
M_{\alpha\gamma}({\bf Q}) \tilde C_{\gamma\beta}({\bf Q}) + \tilde C_{\alpha\gamma}({\bf Q}) M^T_{\gamma\beta}({\bf Q}) = 2R_{\alpha\beta}({\bf Q}).
\end{equation}
One now has to solve the above linear equation with 3 variables to obtain the corresponding correlation functions. Note that the same formalism could be applied
to systems having more than one charge type; the corresponding equations, while remaining linear, will increase substantially in complexity. The solution for the steady state correlation functions is
\begin{widetext}
\begin{align}
&\tilde C_{11}({\bf Q}) = 
\frac{2Q \bar{\rho}_1\big( D_1(\beta_1/\beta_2-1)m_1 m_2 e^{-2QL} + (2Q+m_2)(2(D_1+D_2)Q + D_1 m_1 + D_2 m_2) \big)}{(2(D_1+D_2)Q+D_1 m_1 + D_2 m_2)((2Q+m_1)(2Q+m_2) - m_1 m_2 e^{-2QL})}\label{cfs1}
\\
&\tilde C_{12}({\bf Q}) = \tilde C_{21}({\bf Q}) =
-\frac{2q_1 q_2 \bar{\rho}_1 \bar{\rho}_2 Q \, e^{-QL} (\beta_1 D_1 (2Q+m_1) + \beta_2 D_2 (2Q+m_2))}{\epsilon (2(D_1+D_2)Q+D_1 m_1 + D_2 m_2)((2Q+m_1)(2Q+m_2) - m_1 m_2 e^{-2QL})}\label{cfs2}
\\
&\tilde C_{22}({\bf Q}) = 
\frac{2Q \bar{\rho}_2\big( D_2(\beta_2/\beta_1-1)m_1 m_2 e^{-2QL} + (2Q+m_1)(2(D_1+D_2)Q + D_1 m_1 + D_2 m_2) \big)}{(2(D_1+D_2)Q+D_1 m_1 + D_2 m_2)((2Q+m_1)(2Q+m_2) - m_1 m_2 e^{-2QL})}\label{cfs3}
\end{align}
\end{widetext}
\begin{floatequation}
\mbox{\textit{see eqs.~\eqref{cfs1},\eqref{cfs2} and \eqref{cfs3}}}
\end{floatequation}
where $m_\alpha = \beta_\alpha \overline \rho_\alpha q^2_\alpha/\epsilon$ are inverse Debye lengths (for each system confined to a plane).

\section{The average value of the perpendicular force}
From the steady state correlation functions it is now easy to compute the average thermal Casimir force between the plates in the non-equilibrium state considered here, we find that 
\begin{align}
\langle f_\perp(L)\rangle &=\nonumber \\
 &- S \int \!\! \frac{d{\bf Q}}{(2\pi)^2} \frac{Q \,T_1 T_2 \, m_1 m_2 e^{-2QL}}{(m_1+2Q)(m_2+2Q) - m_1 m_2 e^{-2QL}} \nonumber \\
& \times\frac{\beta_1 D_1 m_1  + \beta_2 D_2 m_2 + 2(\beta_1 D_1 + \beta_2 D_2)Q}{D_1 m_1  + D_2 m_2 + 2(D_1 + D_2)Q}.
\label{eq:force_two_temps}
\end{align}
If one sets $T_1=T_2=T$ in the above, the second line of the above (which is part of the integrand) is equal to $\beta=1/T$ and we immediately recover the equilibrium result  $\langle f^{(eq)}_{\perp}(L,T_1)\rangle$ found in \cite{dea14}. Also if the two systems are at different temperatures but are {\em materially identical} in that $D_1=D_2$ and $m_1=m_2$ we see that
\begin{equation}
\langle f_\perp(L,T_1,T_2))\rangle = {\langle f^{(eq)}_{\perp}(L,T_1)\rangle + \langle f^{(eq)}_{\perp}(L,T_2)\rangle \over 2},
\end{equation}
this is a relation first noticed in \cite{dor98} for planar geometries and subsequently shown to be general within the Rytov formulation between identical dielectric bodies held at different temperatures \cite{ant08}.  

Of particular interest in this model is the far field limit, defined by $L \gg m_1^{-1}, m_2^{-1}$, where in the equilibrium case one recovers the universal thermal Casimir force between ideal conductors \cite{jan05,dea14}. For large $L$ the integral in eq.~(\ref{eq:force_two_temps})
is dominated by small $Q$ and one finds
\begin{equation}
\langle f_\perp(L)\rangle \approx - \frac{S \zeta(3)}{8\pi L^3}\times
\frac{T_2 D_1 m_1  + T_1 D_2 m_2}{D_1 m_1  + D_2 m_2},\label{fperplon}
\end{equation}
where $\zeta(z)$ denotes the Riemann zeta function. When $T_1=T_2=T$ the above becomes independent of $D_\alpha$ and  $m_\alpha$ and we find a long distance force which is universal and independent of the microscopic details of the two planes. However once the temperatures are different, even the long distance force depends on the variables $v_\alpha = D_\alpha m_\alpha$ which have the dimensions of velocity. The form of eq.~(\ref{fperplon}) is remarkably simple, but its physical interpretation is far from being obvious. We also notice that if we take $v_1\gg v_2$ by taking a very large diffusion constant $D_1$ or large $q_1$ or $\overline\rho_1$ the force becomes that of a system with a single temperature $T_2$ at equilibrium. In fact  the plane $1$ behaves like a perfect conductor (due to its ability to react instantaneously to an applied electric field) and the thermal Casimir force is independent of its temperature. The general behavior of eq.~(\ref{fperplon}) is of the same form found for the thermal component of the Casimir force in \cite{dor98,ant05,ant08} where a dielectric system was considered. Note that in this classical model there are no constant bulk contributions to the Casimir force. This is because here the only electrical fields are generated by the charges composing the two planes in the systems whereas in a full quantum calculation there are additional contributions to the electromagnetic field coming from blackbody radiation.

In the near field limit defined by $L \ll m_1^{-1}, m_2^{-1}$ the dominant contribution in eq.~(\ref{eq:force_two_temps}) comes from modes with wave-vectors of amplitude $Q \sim L^{-1}$ and here we obtain
\begin{equation}
\langle f_\perp(L)\rangle \approx -S\left( \frac{T_2 D_1 + T_1 D_2}{D_1 + D_2} \right) \frac{\,m_1 m_2}{16\pi L}. 
\label{eq:f_perp_two_temps}
\end{equation}
In the case where $T_1=T_2=T$ we recover the equilibrium near-field result. However we see here that the way in which the overall force is weighted with the two temperatures depends on the diffusion constants $D_\alpha$, in contrast to the far field limit where it is determined by the velocities $v_\alpha$. The results for the average force in the system with two plates at different temperature show that the force in both the near and far field regimes keep the same functional dependence on $L$ and that the extent of  both regimes is still determined by the relative values of $L$ and $m_\alpha^{-1}$. The weighting of the contribution of each  temperature of the force however depends on the dynamical variables
$D_\alpha$ and $v_\alpha$. We note that the full temperature dependence for this system  is fully determined by the dependence of $m_\alpha$ on the temperature (which is known), as well as the dependence of $D_\alpha$ on the 
temperature of the plate $\alpha$.  
\section{Force fluctuations}
\begin{figure}
		\includegraphics[width=0.45\textwidth]{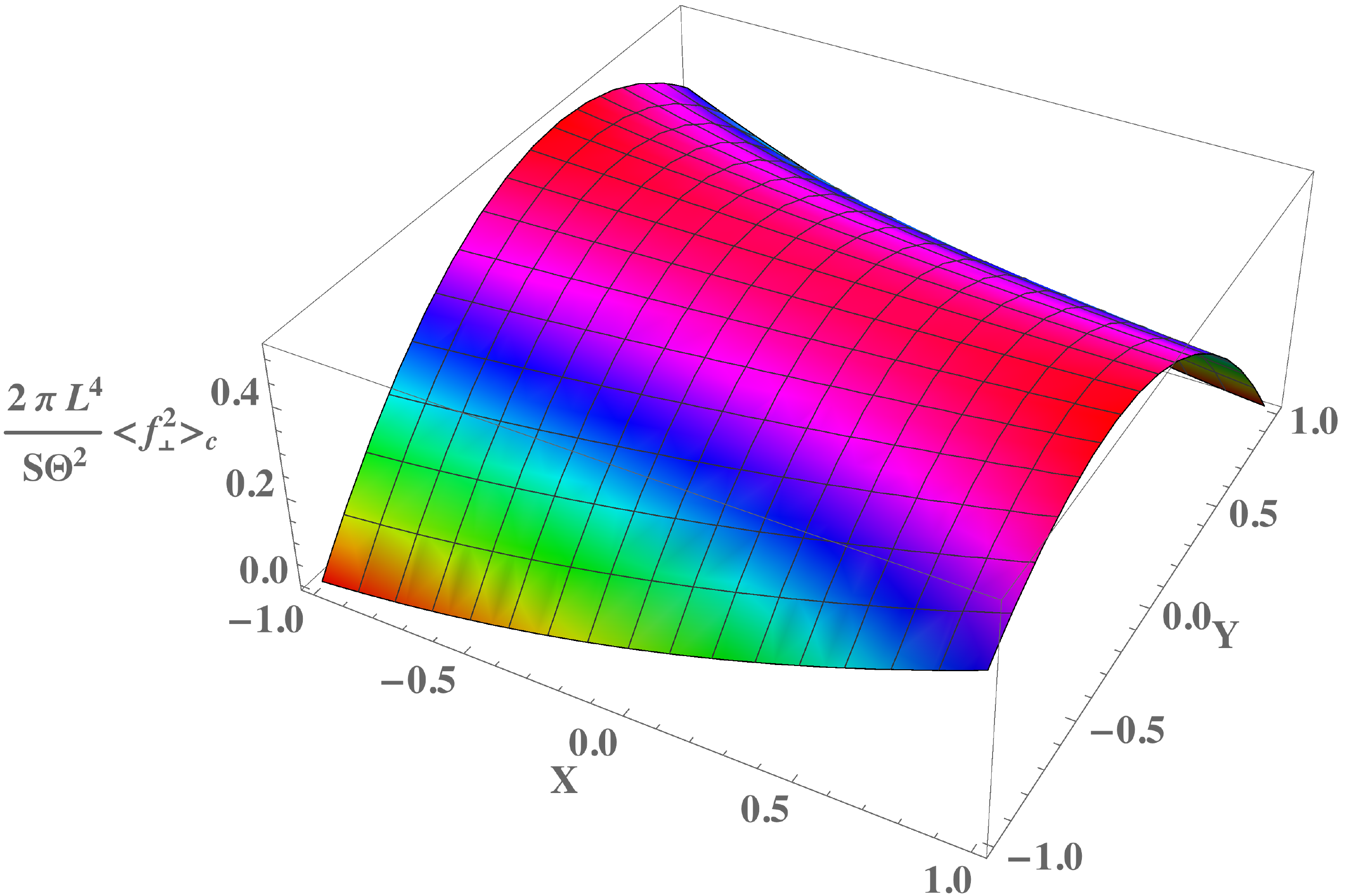}
	\caption{Behavior of the variance of the perpendicular thermal Casimir force in the far field limit, as a function of $X \equiv (D_1 m_1 - D_2 m_2)/(D_1 m_1 + D_2 m_2)$ and $Y \equiv (T_1 - T_2)/(T_1 + T_2)$.}
\label{fig:force_variance_perp}
\end{figure}
\begin{figure}
		\includegraphics[width=0.48\textwidth]{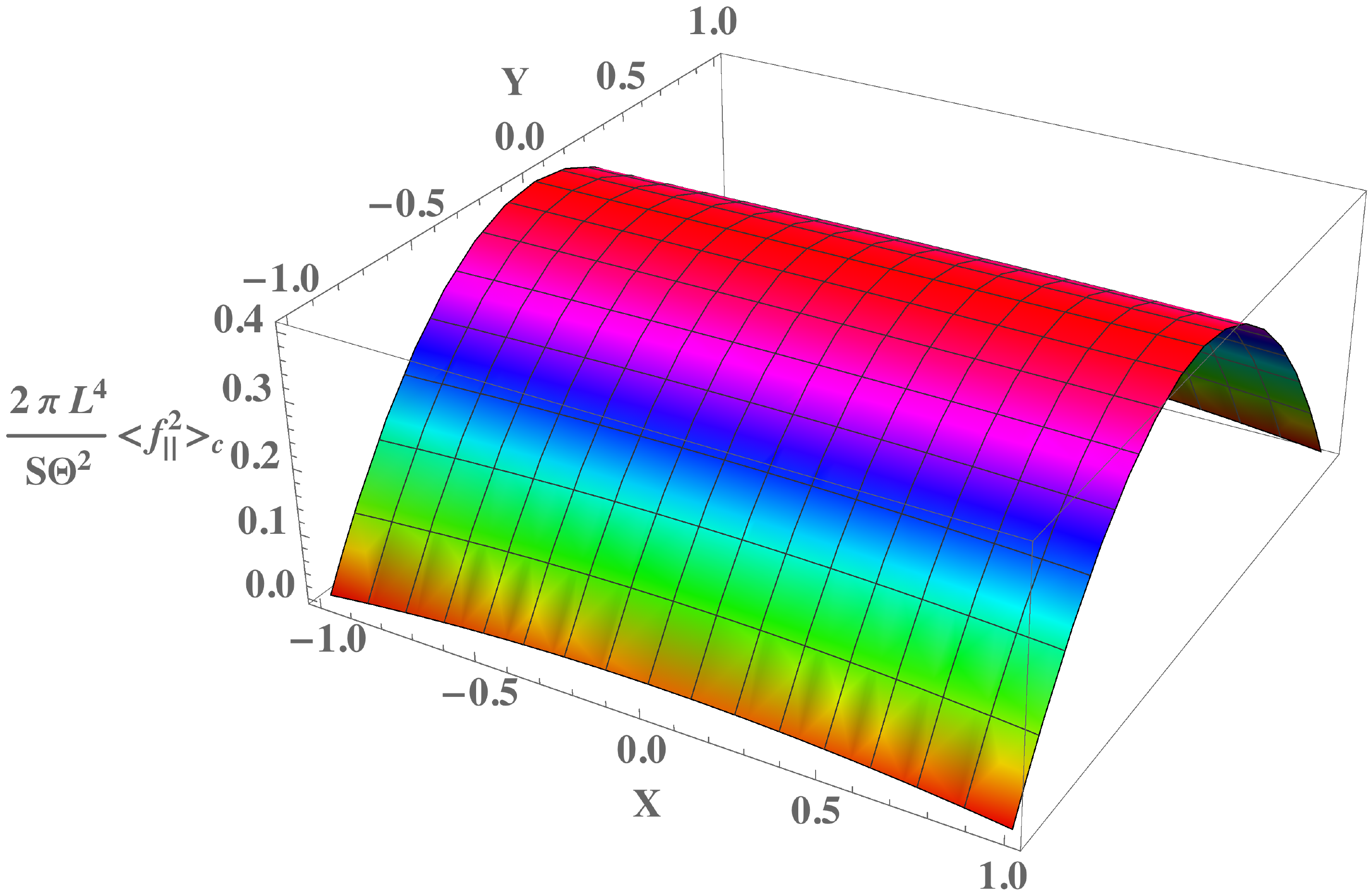}
	\caption{Behavior of the variance of the lateral thermal Casimir force in the far field limit, as a function of $X \equiv (D_1 m_1 - D_2 m_2)/(D_1 m_1 + D_2 m_2)$ and $Y \equiv (T_1 - T_2)/(T_1 + T_2)$.}
\label{fig:force_variance_lateral}
\end{figure}
As both the perpendicular force eq.~(\ref{fp}) and lateral force
eq.~(\ref{fl}) are given in terms of the density fluctuations, we can see that 
fluctuations of the densities will induce the fluctuations of the forces. 

Using Wick's theorem 
we find that the variance of the perpendicular force is given by
\begin{align}
&\langle f_\perp^2\rangle_c = S \, {q_1^2 q_2^2\over \epsilon^2}  \nonumber \\
&\int\!\frac{d{\bf Q}}{(2\pi)^2}\left[\tilde C_{11}(-{\bf Q}) \, \tilde C_{22}({\bf Q}) 
\frac{\partial \tilde G({\bf Q}, L)}{\partial L} \frac{\partial \tilde G(-{\bf Q}, L)}{\partial L}\right.+\nonumber \\
&\left.\tilde C_{12}(-{\bf Q}) \, \tilde C_{21}({\bf Q})
\left( \frac{\partial \tilde G({\bf Q}, L)}{\partial L} \right)^2\right].
\end{align}
Substituting in the correlation function into this expression then yields
\begin{align}
&\langle f_\perp^2\rangle_c \!=\!
S \!\!\int\!\!\!\frac{d{\bf Q}}{(2\pi)^2} \!
\left[
\frac{T_1 T_2 \, m_1 m_2 \, Q^2 \, e^{-2QL}}
{((2Q+m_1)(2Q+m_2) - m_1 m_2 e^{-2QL})^2} \right.
\nonumber
\\
&\qquad\qquad\times
\left( 2Q + m_1 - \frac{D_2(1-\beta_2/\beta_1)m_1 m_2 e^{-2QL}}{(2Q+m_1)D_1+(2Q+m_2)D_2} \right)
\nonumber\\
&\qquad\qquad\times
\left( 2Q + m_2 - \frac{D_1(1-\beta_1/\beta_2)m_1 m_2 e^{-2QL}}{(2Q+m_1)D_1+(2Q+m_2)D_2} \right)
\nonumber\\
&\qquad\quad+\frac{m_1^2 \, m_2^2 \, Q^2 \, e^{-4QL}}{((2Q+m_1)(2Q+m_2) - m_1 m_2 e^{-2QL})^2}
\nonumber\\
&\qquad\qquad\times
\left.
\left( \frac{T_2 D_1 (2Q+m_1) + T_1 D_2 (2Q+m_2)}{D_1 (2Q+m_1) + D_2 (2Q+m_2)} \right)^2
\right].
\label{eq:force_variance_perp}
\end{align}
The variance of the force fluctuations is thus proportional to the surface $S$ of the plates, this is typical thermodynamic scaling for equilibrium systems and persists in the non-equilibrium case when the plate temperatures are unequal. The existence of such thermodynamic scaling along with the absence of any divergences requiring the introduction of an ultra-violet cut-off, ensures that the variance of the force fluctuations must take the form of eq.~(\ref{fpf}) by dimensional analysis. We note that zero temperature 
studies of the quantum fluctuations of the Casimir force between two conductors require
the introduction of a cut-off corresponding to the averaging of the force over a finite time \cite{bar91a,bar91b,ebe91,jae92}. Studies of the thermal fluctuations of the force between dielectrics  find the same behavior as found here with but with a different prefactor \cite{dea13}. The case of massless free fluctuating fields with Dirichlet boundary conditions also has an $L$ (plate separation) dependent part of the force fluctuation variance which has the same form as eq.~(\ref{fpfe}) \cite{bar02}. In both the quantum case and the thermal field case the force on a single (or two infinitely separated plates) has a non-zero
variance; as mentioned above this is due to the fact that in these studies there is a contribution to the electric field which is independent of the charges in the interacting bodies and is generated by quantum \cite{bar91a,bar91b,ebe91,jae92} or thermal \cite{bar02} fluctuations of the field in the surrounding medium.

The rather complicated expression eq.~(\ref{eq:force_variance_perp}) can be simplified in the near and far-field limits. In the far field limit it is given by
\begin{align}
&\langle f_\perp^2\rangle_c \approx \frac{S\Theta^2}{2\pi L^4}\left[ \frac{3\zeta(3)}{4} -\frac{\pi^4}{240} - \frac{3}{8}Y^2 \right.
\nonumber \\ &
\left.
-\left( \frac{3\zeta(3)}{2} -\frac{\pi^4}{60} \right)XY 
+\left( \frac{3}{8} -\frac{\pi^4}{80} +\frac{3\zeta(3)}{4} \right)X^2Y^2 \right]\label{fpf}
\end{align}
where $\Theta = (T_1+T_2)/2$,  and the temperature and velocity mismatches are defined by $Y = (T_1-T_2)/(T_1+T_2)$ and $X = (v_1-v_2)/(v_1+v_2)$. We thus see that in the far field limit $\langle f_\perp^2\rangle_c/\langle f_\perp\rangle^2 \sim L^2/S$.

The behavior of $\langle f_\perp^2\rangle_c$ is plotted in Fig.~\ref{fig:force_variance_perp}. We see that $\langle f_\perp^2\rangle_c$ is maximum for $X=Y=0$, and decreases to zero at $X=Y=\pm 1$. 
From the above equation we also see that in the equilibrium case $T_1=T_2$ we have
\begin{equation}
\langle {f_\perp^{(eq)}}^2\rangle_c \approx \frac{3ST^2}{8\pi L^4}\left(\zeta(3) -\frac{\pi^4}
{180}\right).\label{fpfe}
\end{equation}
The term proportional to $XY$ above is negative, thus if $X$ and $Y$ have the same sign fluctuations are suppressed. If $a_\alpha$ denotes the ionic radius of the ions in
plate $\alpha$ and $\eta'_\alpha$ the background viscosity, then we see that $v_\alpha = \overline \rho_\alpha q^2_\alpha/6\epsilon \pi \eta'_\alpha a_\alpha$ by applying the Stokes formula for the diffusion constant (assuming hard sphere ions). If the viscosity depends only weakly on the temperature then so does $X$. In this regime changing the relative 
temperature in each plate then either increases or decreases the fluctuations.

We have also seen that in the far field limit, when $v_1=v_2$ the average equilibrium force is the arithmetic mean of two equilibrium systems, one with temperature $T_1$ and the other with temperature $T_2$. In this case $X=0$ and we see that 
\begin{equation}
\langle f^2_\perp\rangle_c \approx \frac{3S\Theta^2}{8\pi L^4}\left(\zeta(3) -\frac{\pi^4}
{180} - {1\over 2}Y^2\right).
\end{equation}
We thus see that in this case, and in particular the case of materially identical plates but with different temperatures, the non-equilibrium system has a force which fluctuates less than the corresponding equilibrium one where each plate has the temperature $\Theta = (T_1+T_2)/2$. 

In the near field limit we find that 
\begin{equation}
\langle f_\perp(L)^2 \rangle_c \approx \frac{S T_1 T_2 m_1 m_2}{32\pi L^2}.
\label{eq:rms_two_T_nearfield}
\end{equation}
Recalling that one must have $L \ll m_\alpha^{-1}$ to be in this regime and that $m_\alpha
=\overline \rho_\alpha \beta_\alpha q_\alpha^2/\epsilon$, we see that the fluctuations 
of the force are independent of the two temperatures in this regime. We also see that the dominant fluctuations of the near-field force are also independent of the diffusion constants of the charges in the two plates. In the near field limit we find that $\langle f_\perp^2\rangle_c/\langle f_\perp\rangle^2 \sim 1/Sm_1m_2$ and so the relative fluctuations of the force with respect to its average value become independent of $L$.

The average value of the lateral force is clearly zero, whereas its connected fluctuation correlator is given by
\begin{align}
&\langle f_{||,i} f_{||,j} \rangle_c 
= 
S \, {q_1^2 q_2^2\over \epsilon^2} \!\! \int \!\! \frac{d{\bf Q}}{(2\pi)^2} 
Q_i Q_j \tilde G({\bf Q},L)
\nonumber \\ 
&\qquad\qquad\quad\times
[\tilde C_{11}(-{\bf Q})\tilde C_{22}({\bf Q}) 
\tilde G(-{\bf Q},L)
\nonumber \\ 
&\qquad\qquad\qquad
-\tilde C_{12}(-{\bf Q})\tilde C_{21}({\bf Q})\tilde G({\bf Q},L)].
\label{eq:force_variance_lateral}
\end{align} 
The calculation of the fluctuations of the lateral force can almost be read of from that for the perpendicular force, as $\partial \tilde G({\bf Q}, L)/{\partial L}= -Q \tilde G({\bf Q}, L)$, and 
one finds in the far-field regime that  
\begin{align}
&\langle  f_{||,i} f_{||,j} \rangle_c 
\approx 
\frac{S\Theta^2}{4\pi L^4}
\left[ 
\frac{\pi^4}{240}
- \frac{3}{8}Y^2
+ \left(\frac{3}{8} -\frac{\pi^4}{240}\right) X^2 Y^2 
\right]\delta_{ij}.
\label{ffpar}
\end{align}
In principle there can be correlations between the lateral and transverse forces, but in the present case  it can be verified there is no  correlation. The $i=j$ contribution gives the variance of the lateral force in an arbitrary direction. The behavior is plotted in fig.~\ref{fig:force_variance_lateral}. We see that $\langle f_{||}^2\rangle_c$ is maximum for the plates in thermal equilibrium ($Y=0$), and decreases to zero at $|X|=|Y|=1$. 
In the equilibrium case we find from eq.~(\ref{ffpar}) that
\begin{equation}
\langle {f_{||}^{(eq)}}^2\rangle_c \approx \frac{ST^2\pi^3}{480 L^4}.
\end{equation}
As is the case for the perpendicular force fluctuations, when $X=0$ we see that the 
effect of a temperature difference is to decrease the fluctuations with respect to that of
an equilibrium system with the same {\em average} temperature. An interesting difference between the lateral force fluctuations and the perpendicular force fluctuations is that the
latter depends on the relative signs of $X$ and $Y$ through its dependence on $XY$ while the former does not. Finally the dominant near field contribution to the lateral force fluctuations is identical to that for the fluctuations of the perpendicular force given in eq.~(\ref{eq:rms_two_T_nearfield}). 

In the case where plate 1 is taken to be a perfect conductor,  and so $X=1$, we see that  in the far field limit
\begin{subequations}
\begin{eqnarray}
&&\langle {f_{\perp}}^2\rangle_c \approx \frac{S T_2}{2\pi L^4} 
\left(  
\frac{3\zeta(3)T_2}{4} +\frac{\pi^4(T_1-T_2)}{240}
\right)
\\
&&\langle {f_{||}}^2\rangle_c \approx \frac{ST_1T_2\pi^3}{480 L^4}.
\end{eqnarray}
\end{subequations}
we see here that while the average perpendicular force in this case in independent of the temperature $T_1$ of the perfect conductor, its variance and that of the lateral force do dependent explicitly on the temperature of the perfectly conducting plate.

\section{Conclusions}
We have studied a simple model of Brownian conductor plates giving rise to a thermal Casimir interaction between the said plates. The model enables us to study the out of equilibrium behavior of thermal Casimir interaction between the plates when both plates have different temperatures. We are able to obtain explicit expressions for the non-equilibrium attraction between the plates as a function of the microscopic parameters of 
the model. In particular the far field interaction shows how the universal Casimir interaction between conductors is modified by a temperature difference. In contrast to the equilibrium case, the non-equilibrium far field force depends explicitly on the dynamics of the charges 
through their respective diffusion constants as well as the {\em static} Debye lengths of 
both sets of charges. A similar dependence is also seen in the near-field force, however in both cases the dependence of the force on $L$ is the same as that found in equilibrium ($1/L^3$ for the far-field and $1/L$ for the near-field). The non-equilibrium modification
can thus be thought of as giving a different non-equilibrium {\em Hamaker} constant in both regimes. The formalism employed has also allowed the computation of the fluctuations of 
both the perpendicular and lateral forces between the plates. 

Again the non-equilibrium 
system has force fluctuations which have the same dependence on $L$ as the equilibrium force, only the temperature dependent prefactors are changed. We have also seen that
when both systems are {\em identical}, in that they have the same Debye lengths and diffusion constants (but note that the Debye length depends on the temperature of the plate), the overall force is given by the arithmetic average of two equilibrium systems at both temperatures (as is generally the case for two identical bodies in the Rytov formalism \cite{ant08}). However, this does not hold for the fluctuations of the force (both perpendicular and lateral), which are smaller between {\em identical} plates 
with different temperatures than the equilibrium case where both have the same 
temperature $(T_1+T_2)/2$. 

An obvious extension of this work would be to address the effects of inertia on the ion dynamics and also to treat the electromagnetic interactions relativistically. Using a stochastic density functional theory exists for partially damped dynamics, involving both the density fields and currents of both species \cite{nak2009},  may render a relativistic calculation feasible. However, the far field results here should be relevant up to some length scale, as the near field regime is determined by the parameters $m^{-1}_\alpha$ which are typically nanometric for electrolyte systems.  The size of the relative relativistic correction to the Coulomb potential generated electric field by a moving charge in one plane in the other can be derived from a low velocity expansion of the Li\'enard-Wiechert potential. This correction the has order of magnitude $R\sim L \dot v/c^2$, where $v$ is the particle velocity and $c$ the speed of light. However for a Brownian particle, both the velocity $v$ and acceleration $\dot v$ have correlation functions which diverge at equal times. There must however be a microscopic time scale $\tau$ below over which the velocity $\tau$ is correlated (the ballistic regime). Using this, to regularize the calculation of the variance of $R$,  gives $R\sim L \sqrt{D/\tau^3}/c^2$. Applying the Stokes Einstein for the ionic diffusion constant $D$ and the estimates  $\tau =  a^2 \rho_f /\eta$ \cite{hin1975}, where $a$ is the ionic radius and $\rho_f$ is the volumic  mass density, then suggests that retardation effects are negligible up to length scales of the order of a meter for nanometric ions in water.

\end{document}